\documentclass[pra,aps,twocolumn,floats]{revtex4}
\usepackage{epsfig}
\usepackage{graphicx}% Include figure files
\usepackage{dcolumn}% Align table columns on decimal point
\usepackage{bm}% bold math

\begin{document}
\title{Electric field induced charge noise in doped silicon: ionization of phosphorus donors}

\author{A. J. Ferguson, V. C. Chan, A. R. Hamilton and R. G. Clark}

\address{Centre for Quantum Computer Technology, University of New South Wales, NSW 2052, Australia }

\begin{abstract}
We report low frequency charge noise measurement on silicon
substrates with different phosphorus doping densities. The
measurements are performed with aluminum single electron transistors
(SETs) at millikelvin temperatures where the substrates are in the
insulating regime. By measuring the SET Coulomb oscillations, we
find a gate voltage dependent charge noise on the more heavily doped
substrate. This charge noise, which is seen to have a 1/f spectrum,
is attributed to the electric field induced tunneling of electrons
from their phosphorus donor potentials.
\end{abstract}

\date{\today}

\maketitle
\newpage
With their high charge sensitivity single electron transistors
(SETs) provide a valuable tool for investigating charge transport at
the single electron level. These sensitive electrometers have been
employed to directly observe electron traps in Al$_{2}$O$_{3}$
\cite{zor96}, and electron tunneling through GaAs quantum dots
\cite{lu04} amongst other systems. With particular relevance to this
work, SETs have been used to study both time and frequency domain
charge noise \cite{kuz89,zim92,sta99,fuj00} in semiconductor
substrates. Charge noise in semiconductors has long been of interest
\cite{dut81}, and recently has become an important issue for
solid-state quantum computation where it is crucial that such noise
is minimized to extend quantum coherence times.

We study the effect of electric fields on the charge noise in Si
substrates with different doping levels. At millkelvin temperature
the substrates are in the insulating regime
$(n<n_{MIT}=3.45\times10^{18}cm^{-3})$. For the intentionally doped
substrate $(n=8\times10^{16}cm^{-3})$ we find an additional 1/f
charge noise on application of an electric field. By contrast, this
charge noise was not measurable in the nominally intrinsic sample
$(n=1\times10^{14}cm^{-3})$. We therefore infer that its origin is
electrons tunneling from their phosphorus donors \cite{mar04,smi03}.
While this ionization effect can occur for both substrates, in the
intentionally doped material a larger and more easily measurable
charge redistribution is present due to the higher doping density.

Dopant ionization has previously been measured by transient currents
due to the build-up of depletion regions in p-n junctions
\cite{ros84, dar95}, the slow formation of accumulation layers in
MOS capacitors \cite{goe67} and hysteretic effects in MOSFETs
\cite{gho90}. These experiments were mostly performed at an
intermediate temperature range $(T=4K-20K)$ where ionization can be
thermally activated by the Poole-Frenkel effect \cite{fot90}. At our
experimental temperatures $(T\sim100mK)$, electric field assisted
ionization is no longer dominated by thermal activation but is
expected to be purely field assisted tunneling.

We fabricate the SETs using a standard bilayer resist and a shadow
evaporation process. The electrostatic gate structure is evaporated
Ti/Au and patterned in a separate electron beam lithography step
(see fig. 1b inset). Both silicon substrates are terminated by 5nm
of SiO$_{2}$ grown by thermal oxidization.

\begin{figure} [h]
\begin{center}
\epsfxsize=8cm \epsfbox{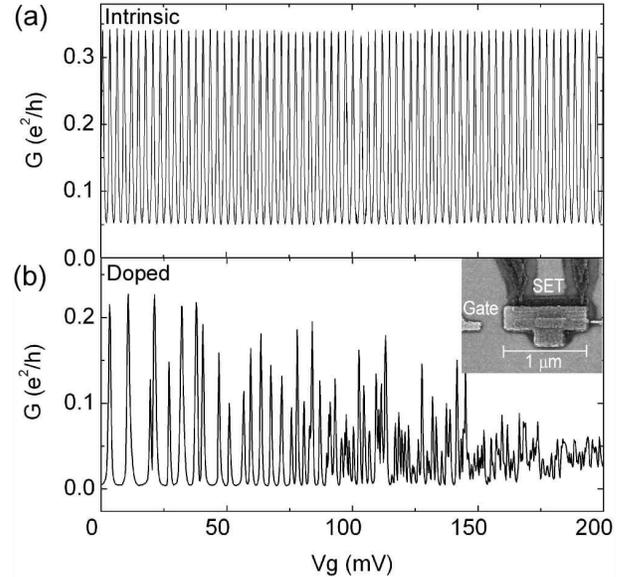}
\end{center}
\caption{a) Coulomb oscillations in the SET on the intrinsic
substrate, showing no deviation from expected behavior. b) An
identical measurement on the highly doped substrate. Contraction of
the oscillation period and subsequent decay of the oscillations into
noise is observed with increasing gate voltage.}
\end{figure}

The SET conductance was measured using standard lock-in techniques
at an excitation frequency of 1kHz and an amplitude of $20\mu V$.
For the noise measurements a FFT analyzer was connected to the
in-phase output of the lock in amplifier (LIA). Typically the range
0.244 Hz to 97.6 Hz was studied with a resolution bandwidth of 0.244
Hz. The sample was mounted in a dilution refrigerator with a base
temperature of approximately 100mK and a 1T magnetic field was
applied to suppress superconductivity in the SETs.

Measurements of the SET fabricated on the intrinsic material (fig.
1a) show periodic Coulomb oscillations as a function of gate
voltage. Little deviation from the periodicity is noticed over the
voltage range swept and, in addition, there is no obvious change in
the noise on the oscillations.

In the case of the doped substrate (fig. 1b), a significant
departure from periodic charging occurs. Close to zero bias Coulomb
oscillations are observed, then as the gate voltage is increased
beyond 50mV, there is a contraction of the oscillation period and a
reduction in oscillation visibility. Finally above some gate
voltage, typically 150mV, the oscillations are completely suppressed
and a constant, but noisy, conductance is measured. A similar effect
is also experimentally observed for negative gate biases. This
behavior is seen for a number of different devices made on this
doped substrate, so rather than being device specific it is a
general property of the material.

The gate voltage dependent charge noise increases with substrate
doping density suggesting that the phosphorus donors are the source
of the noise.
\begin{figure} [h]
\begin{center}
\epsfxsize=7.5cm \epsfbox{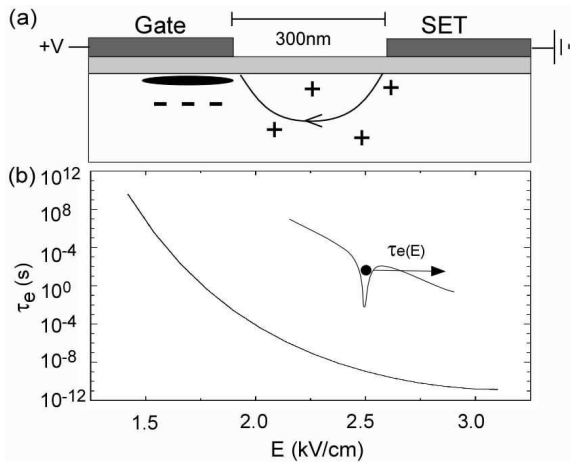}
\end{center}
\caption{a) For a positive gate voltage electrons are removed from
the region of the SET and build up a layer of charge under the gate.
b) The escape time for electrons bound to a Si:P donor on
application of an electric field. This is calculated using a WKB
approximation on a parabolic potential well. Inset showing a
schematic of the tilted potential well.}
\end{figure}
Application of a gate voltage generates an electric field between
the gate electrode and the SET defining the ground plane. It is
known that if the electric field across a dopant is sufficiently
large, electrons (holes) can tunnel from their confining potentials
into the conduction (valance) band (fig. 2a inset). Following the
method of a simple 1-D WKB approximation used to estimate the
tunneling rate of holes from acceptors, we derive an estimate of the
electron emission rate (fig. 2b) from phosphorus donors in an
electric field \cite{fot90}.

For this device in which there is a 300nm gap between the gate and
SET, electric fields in the region of 3kVcm$^{-1}$ are achieved for
an applied voltage of 100mV. At this value of electric field the
electron lifetime is of the order 1ns. It can be seen that the
electric fields applied during these measurements are sufficient to
cause significant donor ionization.

The electric field is highest near the end of the gates and SET.
Ionization therefore happens most rapidly in these regions. However,
the electrons are mobile while the positively charged donors are
not. Taking the gate voltage polarity as in fig 1., electrons that
tunnel into the conduction band are accelerated towards the gate
region where recombination with ionized donors is possible.

The net result is positively charged donors in the SET region and a
layer of electronic charge at the Si-SiO$_{2}$ interface in the
vicinity of the gate (fig. 2a). These positive charges are closer to
the SET than the ionized electrons hence the SET sees a net
additional positive potential. For low total ionization rates, this
positive potential acts as an extra gate voltage. If this ionization
rate increases with electric field then a period contraction will be
observed. The complete loss of contrast in the Coulomb oscillations
is consistent with a high level of charge noise.

Since P donors gradually ionize on application of an electric field,
it is to be expected that the ionization noise is a non-equilibrium
effect and has a time dependence. In order to investigate this time
dependence, the charge noise was quantitatively studied using an FFT
analyzer. In the measurement (fig. 3), the gate bias was ramped from
zero to 300 mV over 30 mins, a gate voltage high enough to produce
significant charge noise. This potential was maintained for 90 mins
and finally ramped down to zero over 30 mins. At regular time
intervals the noise spectrum between 0.244 Hz and 97.6 Hz was
measured with the FFT analyzer. The lowest frequency measured here
(0.24 Hz) corresponds to a time interval of ~4s, and since 8
averages were taken this corresponds to a time of 32s. Since the
noise doesn't change significantly over the measurement time, the
noise measurement approximates one made on a stationary system.

Due to the magnitude of the charge noise experienced by the SET
during these measurements it was not possible to keep the SET biased
at the point of maximum charge sensitivity (halfway up a Coulomb
oscillation). This accounts for the change in contrast between
successive measurements.

As the gate voltage is increased from 0 mV to 300 mV a gradual
increase in charge noise level occurs (fig. 3b). This is consistent
with the observations of figure 1 and is because, at higher gate
biases, more donors experience an electric field sufficient to cause
ionization.
\begin{figure} [h]
\begin{center}
\epsfxsize=7.5cm \epsfbox{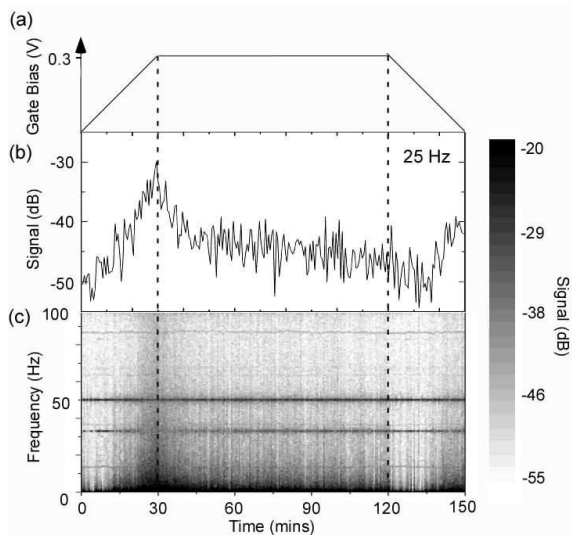}
\end{center}
\caption{a) Schematic of the voltage applied to the SET gate as a
function of time. b) Charge noise measured in a bandwidth of 0.244
Hz around 25 Hz, showing the increase and subsequent equilibration
of this noise. c) Charge noise for this voltage sequence applied to
the highly doped substrate. The horizontal lines are due to mains
noise.}
\end{figure}
In the second phase of the measurement, the gate potential is held
constant for 90 minutes. Over this time there is reduction of the
noise level (fig. 3c) but no return to the initial condition,
presumably due to the long equilibration time at these low
temperatures. A particular donor can ionize only once, so as the
donors gradually ionize fewer events occur which can contribute to
the charge noise. Eventually, on timescales longer than the present
experiment, all available donors will have ionized and the charge
noise returned to its initial value.

Finally, the gate voltage is lowered and almost immediately a
reduction is observed in the noise level. This shows that the system
did not reach equilibrium and can be explained by the sensitivity of
tunneling rate on electric field. When the gate potential is reduced
the electric field across the donors falls, and the escape rates for
electrons still on donor sites decreases causing a sudden drop in
charge noise in the substrate. Before the gate voltage returns to
zero bias, a further onset of charge noise occurs. This is likely to
be recombination noise of the positively charged donors with the
free electrons.

Donor ionization is a random, most likely Poissionian, process with
a rate determined by the the electric field lowered tunnel barrier.
The usual 1/f charge noise in semiconductors is often explained by
an ensemble of two level fluctuators (TLF) having Lorentzian noise
\cite{mac54} distributions with different times constants. Donor
ionization is a similar tunneling process to that in a TLF and hence
also should yield a Lorentzian noise spectrum. In conventional noise
models summing Lorentzians with different time constants gives a 1/f
spectrum \cite{dut81,rog84} and hence a 1/f form would be expected
here.

We measure noise spectra for different biasing conditions (fig. 4).
Spectra are shown for biasing both in the trough and on the side of
a Coulomb oscillation. The differences in noise magnitude here
highlight the changing SET charge sensitivity \cite{sta99}. Also
shown is a measurement taken for a bias where ionization is
occurring and the Coulomb oscillations have lost contrast, showing
yet higher charge noise. All these spectra are close to 1/f,
agreeing with the suggested model.

\begin{figure} [h]
\begin{center}
\epsfxsize=7.5cm \epsfbox{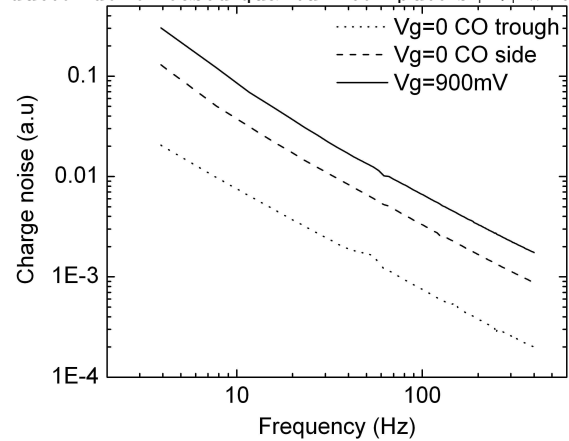}
\end{center}
\caption{Charge noise spectra measured in the absence and presence
of significant ionization. In all cases the form is approximately
1/f. For the spectrum taken in the presence of ionization the
dependence is 1/f$^{1.04}$, a slight deviation is seen at the lower
frequencies.}
\end{figure}

Using the SET as a charge detector we have measured electric field
induced charge noise in insulating silicon substrates at low
temperature. This noise, which depends on the doping density, is due
to the ionization of electrons from their phosphorus donor
potentials. Ionization of donors is of relevance to the noise
properties of a wide range of semiconductor devices including
semiconductor donor based quantum computers \cite{kan98} where it
determines the range of applicable gate voltages.

The authors would like to thank Lloyd Hollenberg and David Reilly
for helpful discussions. This work was supported by the Australian
Research Council, the Australian Government, the U.S. National
Security Agency (NSA), the Advanced Research and Development
Activity (ARDA) and the Army Research Office (ARO) under contract
number DAAD19-01-1-0653.

\bibliographystyle{apsrev}

\newpage
%\begin{description}
%\item [Figure 1]  \item [Figure 2]  \item [Figure 3]

%\end{description}

\end{document}